\documentclass[conference]{IEEEtran}
\usepackage{helvet}
\usepackage{setspace}
\usepackage{titlesec}
\usepackage[letterpaper, portrait, margin=0.75in, top=0.75in, bottom=1.05in]{geometry}
\usepackage{bm} 
\usepackage[noadjust]{cite}

\usepackage{amsmath,amsfonts}

\usepackage[ruled,vlined]{algorithm2e}
\usepackage{amsfonts}
\usepackage{array}
\usepackage{textcomp}
\usepackage{stfloats}
\usepackage{url}
\usepackage{verbatim}
\usepackage{makecell}
\usepackage{graphicx}   % For including images
\usepackage[subrefformat=parens,labelfont=bf,textfont=small]{subcaption} % Modern replacement for subfigure
\usepackage[inkscapelatex=true]{svg}
\def\nb0{{\mathbf{0}}}
\def\nb1{{\mathbf{1}}}

\def\figref#1{Fig.\,\ref{#1}}%

\usepackage{xcolor}
\usepackage{amsmath}
\usepackage{amssymb}
\usepackage{mathtools}
\usepackage{lipsum}

\titlespacing*{\section}{0pt}{1ex}{1ex}  % Adjust space before and after section titles
\titlespacing*{\subsection}{0pt}{0.5ex}{0.5ex}  % Adjust for subsections as well
\IEEEoverridecommandlockouts

\begin{document}
% \bstctlcite{BSTcontrol}
\title{ Impact of Frequency on Diffraction-Aided Wireless Positioning}

\author{~Gaurav~Duggal, ~Anand~M.~Kumar, ~R.~Michael~Buehrer, \\ ~Harpreet~S.~Dhillon, ~Nishith~ Tripathi,  and ~Jeffrey~H.~Reed \\
\thanks{Authors are with Wireless@VT,  Bradley Department of Electrical and Computer Engineering, Virginia Tech,  Blacksburg,
VA, 24061, USA. Email: \{gduggal,anandmk, rbuehrer, hdhillon, nishith, reedjh\}@vt.edu. }
}

% \author{~Gaurav~Duggal \orcid{0000-0002-3765-5479} , 
% ~R.~Michael~Buehrer \orcid{0000-0002-7196-1154},  ~Harpreet~S.~Dhillon \orcid{0000-0003-2829-9449} and ~Jeffrey~H.~Reed \orcid{0000-0003-3494-1901}

% \thanks{G. Duggal, R. M.  Buehrer, H.S. Dhillon and J.H. Reed  are with Wireless@VT,  Bradley Department of Electrical and Computer Engineering, Virginia Tech,  Blacksburg,
% VA, 24061, USA. Email: \{gduggal, rbuehrer, hdhillon, reedjh\}@vt.edu. NIST PSCR PIAP through grant: 70NANB22H070 is gratefully acknowledged.}}

% \vspace{-3mm}
\maketitle
% \vspace{-20mm}

% \author{
% Gaurav Duggal
% \thanks{The authors are with Wireless@VT, Department of ECE, Virginia Tech, Blacksburg, VA. Email: \{gduggal \}@vt.edu. 
% }
% }

% \maketitle

\begin{abstract}
This paper tackles the challenge of accurate positioning in Non-Line-of-Sight (NLoS) environments, with a focus on indoor public safety scenarios where NLoS bias severely impacts localization performance. We explore Diffraction MultiPath Components (MPC) as a critical mechanism for Outdoor-to-Indoor (O2I) signal propagation and its role in positioning. The proposed system comprises of outdoor Uncrewed Aerial Vehicle (UAV) transmitters and indoor receivers that require localization. To facilitate diffraction-based positioning, we develop a method to isolate diffraction MPCs at indoor receivers and validate its effectiveness using a ray-tracing-generated dataset, which we have made publicly available. Our evaluation across the FR1, FR2, and FR3 frequency bands within the 5G/6G spectrum confirms the viability of diffraction-based positioning techniques for next-generation wireless networks. 
\end{abstract}

% \vspace{-4mm}
\section{Introduction}
Indoor public safety scenarios like firefighting or mass shooting incidents require precise location tracking of at-risk individuals, including victims and responders coordinating rescue efforts. Previously, Uncrewed Aerial Vehicle (UAV) based Indoor Positioning Systems (IPS) have been proposed as a solution to this problem \cite{duggal2025indoorpositioningpublicsafety,duggal2025diffractionaidedwirelesspositioning,duggalicc24,duggal2023line, harishetal}. These IPSs offer improved resilience to damaged infrastructure leveraging the fact that in case of an emergency, a UAV based wireless network can be quickly deployed close to the affected area. This offers improved indoor coverage \cite{duggal2023line} however, due to the prevalence of Non-Line-of-Sight (NLoS) conditions, the problem of NLoS bias leads to reduced positioning performance. This occurs because the wireless signal does not propagate along the Euclidean path between the transmitter and receiver. The problem of NLoS bias has been tackled in the literature using several approaches \cite{guvenc2009survey,jiacol,vaghefinlossdp, venkatesh2007nlos}. In these methods, NLoS signal paths are modeled mathematically as the Euclidean distance (between the transmitter and receiver) with an added bounded error, which can help improve position estimates to some extent. In more recent work \cite{leitinger2015evaluation, amiri2023indoor}, the authors introduce propagation mechanism information by showing that when NLoS paths result from reflection, Snell's laws can be applied to straighten the path between the transmitter and receiver, allowing for an exact expression of the path length as the Euclidean distance between the reflected transmitter and receiver. Along similar lines, in our previous work \cite{duggal2025diffractionaidedwirelesspositioning, duggalicc24}, we modeled diffraction as a key propagation mechanism and introduced a novel mathematical path length model based on diffraction, distinct from the conventional Euclidean path model. This served as the foundation for developing multiple positioning techniques tailored for public safety applications. These diffraction-based techniques demonstrated superior performance compared to prior positioning methods that overlooked propagation mechanism information. This suggests that diffraction could be the dominant propagation mechanism in Outdoor-to-Indoor (O2I) scenarios with NLoS signal propagation. However, this naturally raises the following questions: ``How can diffraction paths be isolated amidst other multipath components?" and ``How does the positioning performance of these diffraction-based techniques vary with operating frequency?"
\par
The approach of modeling wireless signal propagation through simple geometric principles originates from Geometrical Optics (GO) and the Geometrical Theory of Diffraction (GTD) / Uniform Theory of Diffraction (UTD) \cite{keller1962geometrical, kouyoumjian1974uniform}, which approximate signal propagation using ray-based models and geometric principles. These ray-based models are well-suited for high frequencies, where reflection and diffraction are predominantly local phenomena influenced by the surface geometry around diffraction and reflection points \cite{balanis2012advanced, namara1990introduction, born2013principles}. 

% This led to the development of a new diffraction-aided positioning technique called Diffraction-Non-linear Least Squares (D-NLS) \cite{duggal2025diffractionaidedwirelesspositioning} suitable for NLoS scenarios and shown to perform better than previous NLoS mitigation positioning techniques.
In this work our main contributions are
\begin{itemize}
    \item \textbf{Positioning Framework}: We present a framework developed from an extensive literature survey to evaluate multipath signal propagation by considering the fundamental propagation mechanisms of Transmission, Reflection, Diffraction, and Diffused Scattering. This framework is designed to assess positioning performance by incorporating mathematical models of electric field strength and path length specific to each propagation mechanism. 
    
    \item \textbf{Diffraction as a propagation mechanism}: In the O2I signal propagation scenario, we show that diffraction from window edges is a dominant propagation mechanism for upper FR1, FR2 and FR3 frequency bands and it is possible to isolate diffraction paths from other signal propagation paths using the First Arriving Path (FAP) principle. Our analysis is based on the positioning framework developed in the first part of the paper where we model the O2I signal propagation scenario in a realistic raytracing software. We have made this dataset and code publicly available \cite{dataset}.
    \item \textbf{Diffraction-aided positioning performance}: We evaluate the 3D positioning performance of the diffraction-aided technique by comparing the Cramér-Rao Lower Bound (CRLB) with its algorithmic implementation, D-NLS, across the FR1, FR2, and FR3 frequency bands. Our analysis demonstrates that diffraction-aided positioning is effective across a broad frequency range, from the upper FR1 band to FR2, and extending to the newly introduced FR3 band.  
\end{itemize}

\section{MultiPath Models}
\label{section_multipath_mathematical_models}
% As the wireless signal propagates in the environment, it interacts with objects in the environment through four physical phenomena - reflection, transmission, diffraction and diffused scattering. Since the transmitted signal reaches the receiver using multiple propagation paths, each path is called a {\em MultiPath Component} (MPC). Previous studies utilizing ToF-based position information linked to each MPC rely on accurate estimation of the corresponding path length \cite{shen2010fundamental}. The accuracy in estimation depends on the mathematical models for path length and the strength of the electric field associated with the MPC.
% Hence, in this section, we present the various mathematical models that can be used to evaluate (a) the associated electric field strength and, (b) the path length corresponding to the various MPCs for each of the propagation mechanisms outlined above. These mathematical models serve as the basis for the ray-tracing simulation software used (Remcom Wireless InSight \cite{wirelessinsight}) that we use to model MPCs generated in an O2I signal propagation scenario. In general, building walls have different constructions, so there are several ways to model them \cite{yasmeen2023estimation}. We assume that the exterior walls of the building are homogeneous and made of concrete, whereas the interior walls are made of plasterboard. 
As wireless signals propagate, they interact with the environment through reflection, transmission, diffraction, and diffused scattering. The transmitted signal reaches the receiver via multiple paths, each termed a {\em MultiPath Component} (MPC). To study wireless signal propagation, previous work has extensively deployed raytracing simulators that use realistic 3D models of the environment in conjunction with computational electromagnetic techniques \cite{yun2015ray}. 
This section introduces a rigorous mathematical framework in which we present models and any underlying approximations with the goal to model (a) the electric field strength and (b) the path length of all MPCs generated due to different propagation mechanisms. This framework is implemented within Remcom Wireless InSight \cite{wirelessinsight}, which we utilize to assess the positioning performance in O2I scenarios. Various techniques have been employed to model building structures, including walls \cite{yasmeen2023estimation}. In this study, we represent the building's exterior as concrete, while the interior structure is partitioned by plasterboard walls. Our objective is to examine the frequency dependence of the diffraction-based positioning technique, which is based on the observation that diffraction MPCs originate from window edge diffraction in an O2I signal propagation scenario. Therefore, this study does not consider indoor propagation effects from furniture and human blockages, as they are not the primary focus and can be addressed using existing techniques.

\subsection{Reflection And Transmission}
\label{section_reflection_and_transmission}
When an incident electromagnetic wave encounters the boundary between two mediums, part of the energy is transmitted, and part is reflected, producing transmitted and reflected rays. The energy distribution between these rays is determined by the Fresnel coefficients, which depend on the material properties, the incident angle, and the field's polarization. Previously, the exterior walls of the building have been modeled as a single-layer lossy non-magnetic dielectric slab \cite{zhekov2020dielectric, richalot2000electromagnetic}. 
\begin{figure*}[htbp]
    \begin{subfigure}{0.31\linewidth}
        \centering
        \includegraphics[clip, trim=0cm 0cm 0cm 0cm, width=1\linewidth]{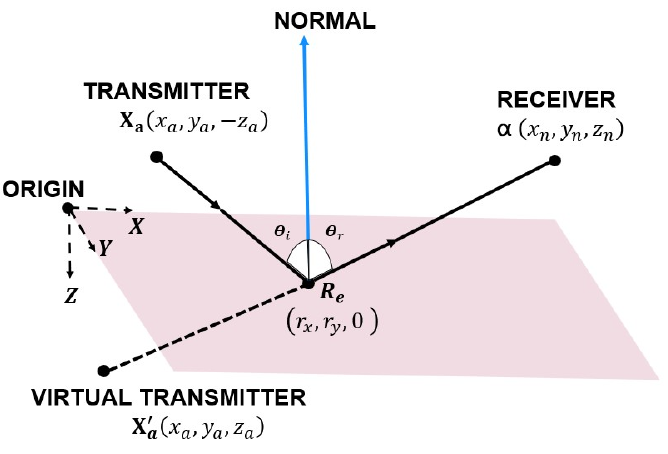}
    \caption{Reflection MPC}
        \label{fig_reflection_mpc}
    \end{subfigure}
    \begin{subfigure}{0.35\linewidth}
        \centering
\includegraphics[clip, trim=0cm 0cm 0cm 0cm, width=1\linewidth]{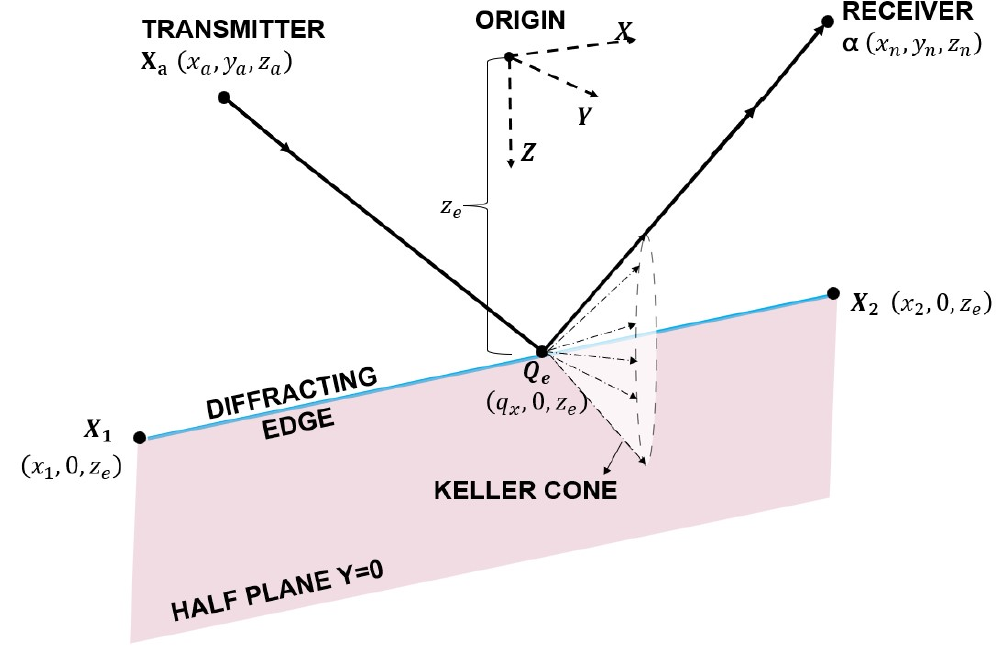}
        \caption{Diffraction MPC}
        \label{fig_diffraction_mpc}
    \end{subfigure}
    \begin{subfigure}{0.31\linewidth}
        % \centering
\includegraphics[clip, trim=0cm 0cm 0cm 0cm, width=1\linewidth]{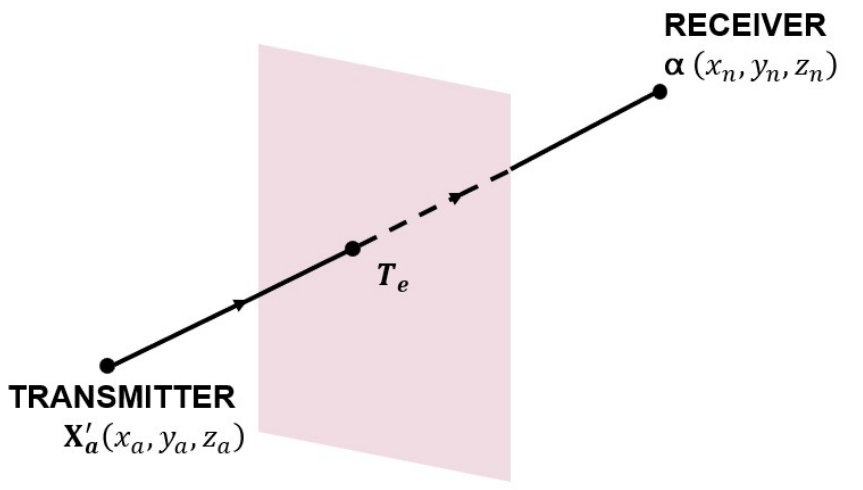}
        \caption{Transmission MPC}
        \label{fig_transmission_mpc}
    \end{subfigure}

\caption{(a) The Reflection MPC path length $\bm{X_aR_e\alpha}$ can be reduced to the Euclidean distance between the position of the virtual transmitter obtained using Snell's law of reflection and the receiver position $\bm{\alpha}$. (b) The path length of the diffraction MPC $\bm{X_aQ_e\alpha}$ is the sum of two Euclidean distances given by \eqref{eq_exact_path_length_diffraction}. (c) Finally, the transmission MPC path length is the Euclidean distance $\bm{X_a\alpha}$.}
    \label{fig_prop_mech}
\end{figure*}

\subsubsection{Path Length Model}
For a transmitter at $\bm{X_a}=[x_a,y_a,z_a]^T$ and receiver at $\bm{\alpha}=[x_n,y_n,z_n]^T$, if there exists a MPC with no obstacles, the associated path length is the Euclidean distance between the transmitter and receiver.% \begin{equation}
% p = \sqrt{(x_a-x_n)^2+(y_a-y_n)^2+(z_a-z_n)^2)}.
% \label{eq_euclidean_path_length}
% \end{equation}
We place a dielectric slab between the Tx-Rx pair to generate a transmission MPC (\figref{fig_transmission_mpc}). Assuming the slab's thickness is negligible relative to the signal's free-space path length, the path length remains unchanged, and refraction effects are ignored.

For reflection modeling, a reflection MPC is considered between the transmitter and receiver (\figref{fig_reflection_mpc}), representing specular reflection from a smooth planar surface. This results in a path length longer than the direct Euclidean distance between the transmitter $\bm{X_a}$ and receiver $\bm{\alpha}$. Using Snell's law ($\theta_i = \theta_r$), the reflection path length is calculated by reflecting the transmitter across the surface to form a {\em virtual transmitter}, creating a Euclidean path to the receiver. 
% This approach has been utilized for leveraging reflection MPCs for NLoS indoor positioning, by {\em Leitinger et al.}\cite{leitinger2015evaluation}.

\subsubsection{Electric Field Strength}
Building materials are modeled as dielectric slabs, where transmission and reflection from these slabs cause attenuation of the incident wireless signal. To calculate the path loss due to both, we use the relative permittivity, $\varepsilon_r'$ and electrical conductivity, $\sigma$ of the dielectric slabs. The path loss $PL_T$ (in $\mathrm{dB/m}$) due to transmissions through a dielectric slab \cite{balanis2012advanced} can be expressed as

\begin{align}
   PL_T &= 12.27 \pi f t \sqrt{\mu_0 \varepsilon_0\varepsilon_r'}
\left(
    \sqrt{1 + \left(\frac{\sigma}{2 \pi f \varepsilon_0\varepsilon_r'}\right)^2}
    - 1
\right)^{0.5}.
\end{align}
Here, $t$ is the thickness of the dielectric slab, $f$ is the frequency of the signal, $\varepsilon_0$ is the permittivity of free space and $\mu_0$ is the permeability of free space. ITU \cite{ITU-R_P2040-1}, presents the relative permittivity and electrical conductivity of various building materials as 
\begin{equation}
\begin{split}
    \varepsilon_r' = af_{GHz}^b, \;\;\;
    \sigma = cf_{GHz}^d.  
\end{split}
\end{equation}
The parameters $a,b,c$, and $d$ for common building materials can be found in \cite{ITU-R_P2040-1}. In general $b=0$ for concrete and plasterboard hence electrical conductivity varies with transmission frequency, whereas the dielectric constant remains independent of it. 

% \begin{table}[h!]
%     \centering
%     \begin{tabular}{|c|c|c|c|c|}
%         \hline
%         \textbf{Material Class} & \textbf{a} & \textbf{b} & \textbf{c} & \textbf{d} \\ \hline
%         Vacuum/Air & 1 & 0 & 0 & 0 \\ \hline
%         Concrete & 5.31 & 0 & 0.0326 & 0.8095 \\ \hline
%         Glass & 6.27 & 0 & 0.0043 & 1.1925  \\ \hline
%         Plasterboard & 2.94 & 0 & 0.0116 & 0.7076 \\ \hline   
%     \end{tabular}
%     \caption{Electrical Properties of Materials}
%     \label{tab:ElectricalProperties}
% \end{table}
% Now we can use the attenuation constant along with Friis Equation to estimate the path loss for the transmission path between a transmitter located outside and receiver located inside a building using:

% \begin{equation}
%     PL_{dB} = 20log_{10}(d) + 20log_{10}(f) +  20log_{10}\left(\frac{4\pi}{c}\right) - G_{tx} - G_{rx} + 20log_{10}(e)\cdot\alpha\cdot d
%     \label{eq:PLMaterial}
% \end{equation}
% where the last term is the additional path loss due to transmission through medium.
\subsection{Modeling Diffraction}
Diffraction occurs when an electromagnetic wave interacts with an edge, as shown in \figref{fig_diffraction_mpc}. A diffraction MPC originates from the transmitter, with the incident ray striking the diffraction point $\bm{Q_e} = [q_x, 0, z_e]$. This interaction creates a cone of diffracted rays, known as the Keller cone \cite{rahmat2007keller}, centered at $\bm{Q_e}$. From this cone, the ray passing through the receiver location is selected, and together with the incident ray, it defines the diffraction MPC.
\subsubsection{Path Length Model}
Now, the location of the diffraction point on the diffracting edge is a complicated function of the position of the transmitter, the position of the edge and the position of the receiver. The closed form expression for the diffraction path length $\bm{X_aQ_e\alpha}$ was derived in \cite{duggal2025diffractionaidedwirelesspositioning, duggalicc24}. The exact path length $p$ traced out by the geometrical path $\bm{X_aQ_e\alpha}$ followed by the diffraction MPC in Fig.\,\ref{fig_diffraction_mpc} is expressed as 
\begin{equation}
\label{eq_exact_path_length_diffraction}
\begin{split}
p = & \sqrt{(x_{a}-q_x)^2+(y_{a})^2+(z_{a}-z_{e})^2} \\ 
&+  \sqrt{(x_n-q_x)^2+(y_n)^2+(z_e-z_n)^2}.\\
\end{split}
\end{equation}
Here, the coordinates of the diffraction point $\bm{Q}_e=[q_x,0,z_e]^T$, can be expressed as a convex combination of the coordinates of the endpoints of the edge $\bm{X}_1=[x_1,0,z_e]^T$ and $\bm{X}_2=[x_2,0,z_e]^T$ as
\begin{equation}
\begin{split}
\bm{Q}_e & = \lambda \bm{X}_1 + (1-\lambda) \bm{X}_2,\;
\lambda  = \frac{-b \pm \sqrt{b^2-4ac}}{2a},  \\
a &  = (x_1-x_2)^2 \left[(y_n^2-y_{a}^2)+ (z_{n}^2-z_a^2)  +2z_e(z_a-z_{n})\right], \\
b &  = 2(x_1-x_2) \left[(x_2-x_a)\left((z_e-z_{n})^2+y_n^2\right) \right. \\& \left. \;\;\;\;\;\;\;\;\;\;\;\;\;\;\;\;\;\;\;\;\;\;\;\;\;\;\;\;\; -(x_2-x_n)  \left((z_e-z_a)^2+y_a^2\right)\right], \\
c & = (x_2-x_a)^2 \left[(z_e-z_{n})^2+y_n^2\right] \\&\;\;\;\;\;\;\;\;\;\;\;\;\;\;\;\;\;\;\;\;\;\;\;\;\;\;\;\;\;   - (x_2-x_n)^2  \left[(z_e-z_a)^2+y_a^2\right].
\label{eq_quadratic_qe}
\end{split}
\end{equation}
Note, the above path length is exact for diffraction. However, as demonstrated in \cite{duggal2025diffractionaidedwirelesspositioning}, achieving positioning in the O2I scenario with minimal {\em a priori} floor map information requires assuming that the vertical distance between the receiver location $\bm{\alpha}$ and $\bm{Q_e}$ is half the window height $w$. This approximation can be intuitively explained by observing that the active diffraction edge always lies on the window located on the same building floor as the receiver.
Hence, for \figref{fig_diffraction_mpc} we assume \begin{equation}
    z_e = z_n + \frac{w}{2}.
\end{equation}
Here, $w$ is the vertical height of the window. This leads to an approximate diffraction path length expression     
\begin{equation}
\label{eq_approximate_path_length_diffraction}
\begin{split}
p = & \sqrt{(x_{a}-q_x)^2+(y_{a})^2+(z_{a}-z_n-0.5w)^2} \\ 
&+  \sqrt{(x_n-q_x)^2+(y_n)^2+(0.5w)^2}.\\
\end{split}
\end{equation}
We use this approximate diffraction path length model for NLoS positioning in later sections. 

\subsubsection{Electric Field}
The diffraction electric field depends on the orientation of the incident wave relative to the edge, the wave’s polarization, and the material properties of the edge. Analogous to Fresnel reflection coefficients, diffraction coefficients enable the calculation of the diffracted electric field’s magnitude. Significant contributions in this area include those by {\em Kouyoumjian et al.} \cite{kouyoumjian1974uniform} for thin perfectly conducting edges, {\em Burnside et al.} \cite{burnside1983high} for thin dielectric edges, {\em Luebbers et al.} \cite{luebbers1984finite,luebbers1988comparison,luebbers1989heuristic} for finite-conductivity and rough lossy wedges, and {\em Holm et al.} \cite{holm2000new} for non-perfectly conducting edges. The full 3D diffraction electric field equations, which include both horizontal and vertical polarization components, are not presented here due to their complexity; further details can be found in \cite{balanis2012advanced, namara1990introduction, born2013principles, duggal2025diffractionaidedwirelesspositioning}.
% The diffraction electric field depends on the direction of the incident field with respect to the edge, the polarization of the incident field, and the material properties of the edge. Similarly to the Fresnel reflection coefficients, we have the diffraction coefficients that can be used to calculate the magnitude of the diffraction electric field. The development of the diffraction coefficients has seen extensive contributions with {\em Kouyoumjian et al. \cite{kouyoumjian1974uniform}}  who proposed diffraction coefficients for a thin perfectly conducting edge, {\em Burnside et al.} \cite{burnside1983high} described diffraction coefficients for a thin dielectric edge, {\em Luebbers et al.} \cite{luebbers1984finite} described diffraction coefficients for edges with finite conductivity and later for rough lossy wedges \cite{luebbers1988comparison,luebbers1989heuristic}  and finally {\em Holm et al.}\cite{holm2000new} described a heuristic diffraction coefficient for non-perfectly conducting edges. Remcom Wireless InSight \cite{wirelessinsight} incorporates the appropriate diffraction coefficients depending on the modeled scenario in the calculation of the diffraction field. The diffraction electric field equations are complex for 3D and omitted from this study. They include both the horizontal and vertical polarization components and can be referred to in \cite{balanis2012advanced, namara1990introduction, born2013principles, duggal2025diffractionaidedwirelesspositioning}. 

\subsection{Modeling Diffused Scattering}
Diffuse scattering is considered to originate from planar surfaces such as building floors and walls, which are modeled as a collection of infinitesimal elements. Each element is assumed to scatter a fraction of the incident power in all directions, governed by the scattering coefficient \cite{degli2001diffuse,degli2007measurement, jiang2022survey} and the Fresnel reflection coefficient, which itself is dependent on the incident field direction. The scattering coefficient varies based on material properties and surface irregularities. The resulting scattered field can be represented as a scattering radiation lobe, where its orientation determines the scattering model: if the lobe's maximum is directed perpendicular to the surface, it follows the Lambertian model, whereas if it is aligned with the specular reflection direction, it follows the Directive model. In our ray-tracing simulations, we employ the Directive model.

\section{First Arriving Path Aided 3D Positioning}
In ToF-based positioning, the receiver estimates path lengths using signals from multiple transmitters or anchors. Since the received signal comprises multiple paths, the estimated path length corresponds to a specific MPC selected from those present at the location. These estimates are influenced by the selected path, the governing mathematical model, and measurement noise. The estimated path lengths are then used to compute the node's 3D position using various algorithms. This section introduces metrics to analyze the MPCs at a receiver location, outlines a method to select an MPC and estimate its path length, and applies algorithms tailored to two path models to determine the receiver's position.  

% \vspace{-1em}  % Reduce space before section
\subsection{Simulation Setup}
\begin{figure*}[!htbp]
    % Large figure on the left
    \begin{subfigure}{0.54\linewidth}
        \centering
        \includegraphics[clip, trim=0cm 0cm 0cm 0cm, width=1\linewidth]{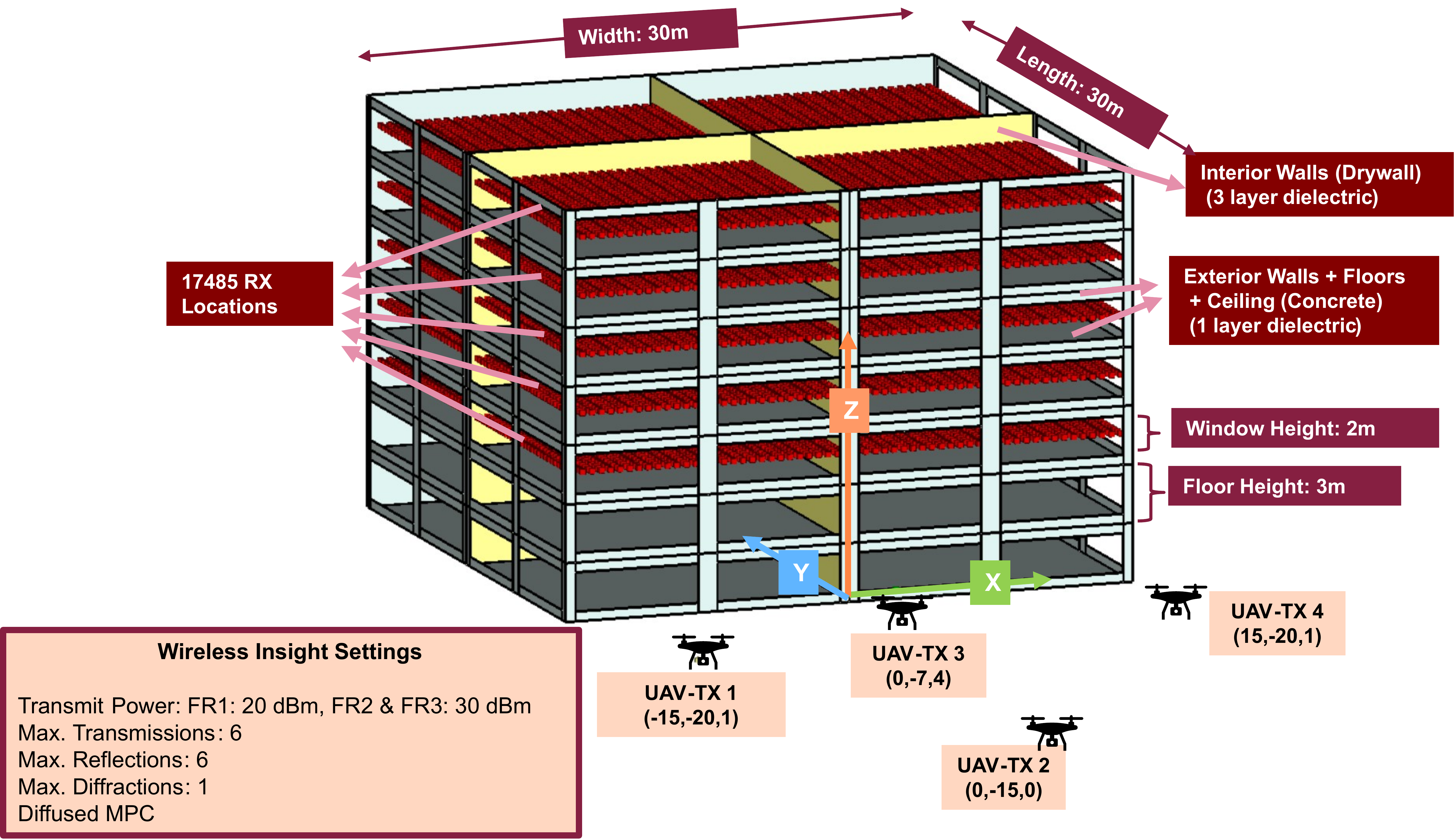}
        \caption{O2I signal propagation scenario}
        \label{fig_O2I_scenario}
    \end{subfigure}
    \hfill
    % Right figure with trimmed white space
    \begin{subfigure}{0.44\linewidth}
        \centering
        \includegraphics[clip, trim=0cm 0cm 0cm 0cm,width=0.8\linewidth]{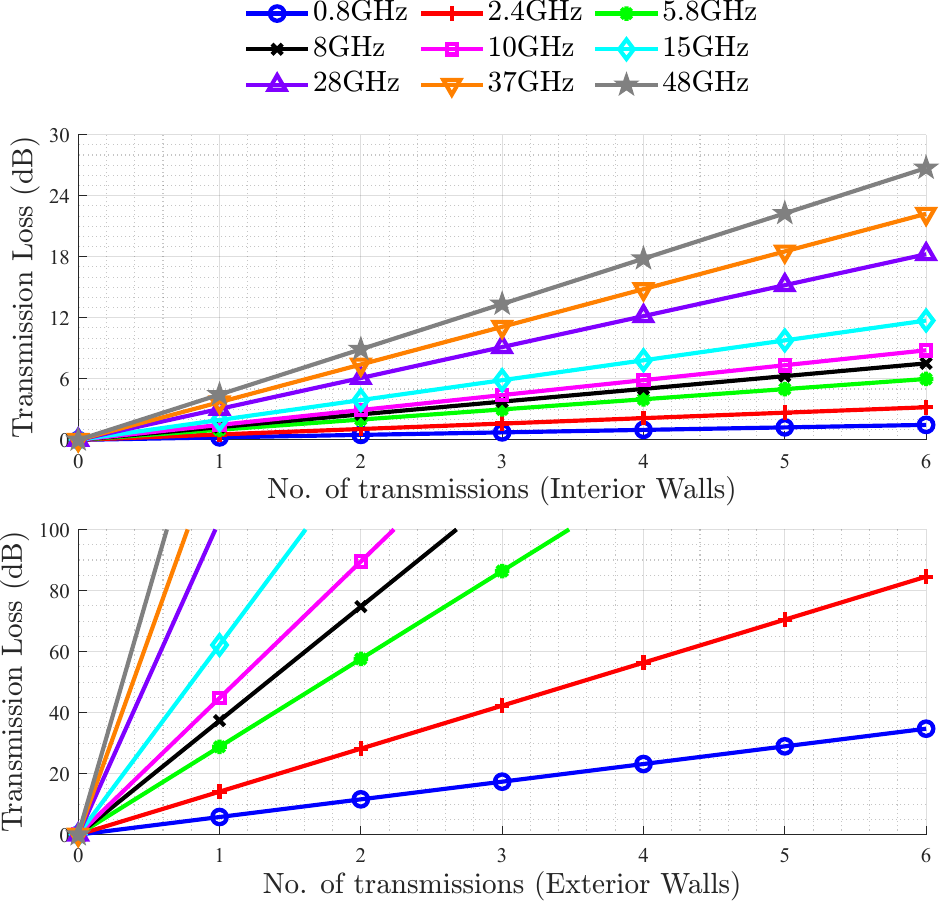}
        \caption{Transmission loss}
        \label{fig_tansmissionLoss}
    \end{subfigure}

    \caption{(a) An O2I signal propagation scenario modeled in our raytracing software involving a building with concrete exterior walls and floors, drywall interiors, and UAV transmitters positioned outside while the receivers are located indoors. (b) The frequency dependent transmission loss through the building materials}
    \label{fig_wireless_insight}
    \vspace{-1.5em}
\end{figure*}
% \vspace{0.5em} 
% Figure \figref{fig_wireless_insight}, describes the building modeled to evaluate NLoS positioning performance. 
Refer to \figref{fig_wireless_insight}, which illustrates a building model with concrete exterior walls, floors, and ceilings represented as 30 cm single-layer dielectric slabs. The interior drywall walls are modeled as a three-layer dielectric, consisting of two 1.3 cm thick plasterboard sheets separated by an 8.9 cm air gap. $4$ UAV transmitters operating with a signal bandwidth $B=400$ MHz are placed outside the building between floors $1$ and $2$. Next, a total of $17,485$ receivers are placed uniformly on a grid, spaced $0.5$m apart, inside the building on floors $3$-$7$. Since our focus is on positioning in NLoS scenarios, receivers were not placed on floors 1 and 2, as these floors predominantly experience LOS transmissions from UAV transmitters through windows. This scenario was created in a raytracing software - Remcom's Wireless InSight \cite{wirelessinsight} to generate the received signal consisting of the 25 MPCs with the highest SNR at each receiver location. Each MPC is simulated to experience up to $6$ transmissions, $6$ reflections, $1$ diffraction and diffuse scattering from all surfaces and edges in the environment. Each MPC has an associated received signal power and ToF calculated using the path length and electric field models described in Section \ref{section_multipath_mathematical_models}.
The SNR for each path is generated using the power of the received signal $P_{rx}$ obtained from the raytracing simulator and the noise floor of the receiver given by $N_0=KTB$, where $K$ is the Boltzmann constant, $T$ is the noise temperature, and $B$ is the signal bandwidth.

\subsection{First Arriving Path Estimation}
\label{section_first_arriving_path_estimator}
The receiver selects the path for ToF estimation using the First Arriving Path (FAP) principle \cite{falsi2006time}.  At each receiver location, the strongest path with an SNR of $S_{\rm MAX}$ is identified by iterating through the Power Delay Profile (PDP). Next, we set a threshold SNR $T$, where $T=S_{\rm MAX}-T_{\rm FAP}$. Here, $T$ is the minimum SNR required for a resolvable path to be chosen and $T_{\rm FAP}$, is a user configurable parameter. Now starting from the strongest path, we look for the shortest path which meets the threshold $T$ upon which we declare it to be the FAP at that location. Hence, increasing $T_{\rm FAP}$ influences the estimator to favor shorter but weaker paths. It has  been shown that the Cramér-Rao Lower Bound (CRLB) to estimate the ToF of a resolvable FAP is \cite{emenonye2024fundamentalsleobasedlocalization} 
\begin{equation}
\label{eq_ranging_CRB}
\begin{split}
    \text{CRLB}(\tau_0) = \lambda= \frac{1}{\sqrt{8\pi^2\beta^2\text{S}_{\tau_0}}}.
\end{split}
\end{equation}
Here, $\beta^2 = \frac{\int_{-\infty}^{\infty} f^2 |X(f)|^2 \, df}{\int_{-\infty}^{\infty} |X(f)|^2 \, df}$ is the mean squared bandwidth of the baseband equivalent of the transmit signal with Fourier transform $X(f)$ and the SNR of the FAP is $\text{S}_{\tau_0}$. 
Hence, our FAP estimation accuracy improves with increasing SNR and Bandwidth.

\subsection{CRLB For Positioning Using The Diffraction Path Model}
\label{section_CRLB}
The CRLB for 3D positioning using the diffraction path model represents a lower bound to the positioning performance achievable in case we are able to obtain a diffraction path length measurement at a given receiver location. In this case the measurement model for the FAP at the receiver location $\bm{\alpha}$ from each of the $M$ UAV transmitters would be
\begin{equation}
\label{eq_measurement_model_diffraction}
\bm{r} = \bm{p}(\bm{\alpha}) + \bm{n}.
\end{equation}
Here, $\bm{r}\in \mathbb{R}^{M\times1}$ is the vector of the estimates of the first arriving paths, $\bm{p}(\bm{\alpha}) =[p_0(\bm{\alpha}),\cdots p_j(\bm{\alpha}),\cdots,p_{M-1}(\bm{\alpha})]^T $ where $p_j(\bm{\alpha})$ is the approximate diffraction path length \eqref{eq_approximate_path_length_diffraction} from the $j^{\text{th}}$ UAV transmitter and $\bm{n}\in\mathbb{R}^{M \times 1}$ is the uncertainty in the range measurements due to receiver noise. The receiver noise is assumed to be independent Gaussian, and the uncertainty in range measurements can be obtained from the CRLB of the ToF of the first arrival path \eqref{eq_ranging_CRB} by scaling it with the speed of light. It can be shown that the Fisher Information Matrix (FIM) $\bm{J}_{\bm{\alpha}}$ for estimating the 3D position $\bm{\alpha}\in \mathbb{R}^{3\times 3}$ using the diffraction MPCs is given by \cite{duggal2025diffractionaidedwirelesspositioning}
\begin{equation}
\begin{split}
    \bm{J}_{\bm{\alpha}} = c^2\bm{J}\bm{J}_{\tau_0} \bm{J}^T.
\end{split}
\end{equation}
Here c is the speed of light and $\bm{J}_{\tau_0} = diag(\lambda_0, \cdots,\lambda_{M-1})$ is a diagonal matrix where $\lambda_j = 8\pi^2\beta^2SNR_j$ represents the inverse of the CRLB of the path delay estimate of the first arriving path between the $j^{th}$ anchor and the node. Finally, $\bm{J} \in \mathbb{R}^{3 \times M}$ is the Jacobian given in \cite{duggal2025diffractionaidedwirelesspositioning}. 
% \begin{equation}
% \label{eq_Jacobian}
% \bm{J} = 
% \begin{bmatrix}
% \frac{\partial{p_{0}(\bm{\alpha})}}{\partial x_n} & \frac{\partial{p_{0}(\bm{\alpha})}}{\partial y_n} &  \frac{\partial{p_{0}(\bm{\alpha})}}{\partial z_{n}}\\
% \vdots& \vdots & \vdots\\
% \frac{\partial{p_{M-1}(\bm{\alpha})}}{\partial x_n} &\frac{\partial{p_{M-1}(\bm{\alpha})}}{\partial y_n} &  \frac{\partial{p_{M-1}(\bm{\alpha})}}{\partial z_{n}}\\
% \end{bmatrix}.
% \end{equation}

\begin{equation}
\label{eq_Jacobian}
\bm{J} = 
\begin{bmatrix}
\frac{\partial{p_{0}(\bm{\alpha})}}{\partial x_n} & \cdots &  \frac{\partial{p_{M-1}(\bm{\alpha})}}{\partial x_n}\\
\frac{\partial{p_{0}(\bm{\alpha})}}{\partial y_n} &\cdots &  \frac{\partial{p_{M-1}(\bm{\alpha})}}{\partial y_{n}}\\
\frac{\partial{p_{0}(\bm{\alpha})}}{\partial z_n} &\cdots &  \frac{\partial{p_{M-1}(\bm{\alpha})}}{\partial z_{n}}
\end{bmatrix}.
\end{equation}

The closed form expression for the elements of the Jacobian are given in \cite{duggal2025diffractionaidedwirelesspositioning}.

The CRLB to the RMSE in the 3D position estimation is thus obtained from the FIM as the Position Error Bound (PEB) \cite{emenonye2024fundamentalsleobasedlocalization} where $\text{PEB}=\sqrt{\text{Tr}({\bm{J_{\alpha}}}^{-1})}$ and $\text{Tr}{(\bm{A})}$ is the trace of matrix $\bm{A}$. 
\subsection{Positioning Algorithms: D-NLS and LLS}
\label{section_LLS_DNLS}
Once we have ranging measurements from $M$ transmitters, we can estimate the 3D position of the indoor receiver using a least squares method. For the diffraction path model, we search for the receiver position $\bm{\alpha}$ that minimizes the residual between the ranging measurements $\bm{r}$ and the diffraction path model $\bm{p}(\bm{\alpha})$ in \eqref{eq_measurement_model_diffraction}. Since the diffraction path model is non-linear, this is achieved using an iterative procedure called \textbf{D-NLS} \cite{duggal2025diffractionaidedwirelesspositioning} as
\begin{equation}
    \bm{\hat{\alpha}}_{m+1} = \bm{\hat{\alpha}}_{m} + (\bm{J}_{m}^T\bm{J}_m)^{-1}\bm{J}_m^T(\bm{r}-\bm{p}(\bm{\hat{\alpha}}_m)). 
\end{equation}
Here, $\bm{\hat{\alpha}}_m$ is the position estimate, $\bm{J}_m$ is the Jacobian of the diffraction path length from M anchors \cite{duggal2025diffractionaidedwirelesspositioning} with the subscript $m$ denoting the iteration index. Linear least squares (LLS) minimizes the residual between the range measurements and the Euclidean path model. The Euclidean path model can be linearized by the squaring operation and this leads to a one-shot non-iterative estimator \cite{zekavat2019handbook}.
\subsection{MPC Analysis Based On Metrics}
For our O2I scenario, for each Rx location within the building, we have several MPCs present. The path followed by a particular MPC can be represented as a string such as ‘Tx-X-X-$\cdots$-X-Rx’. Since the MPC begins at the transmitter and ends at the receiver, all possible strings start with `Tx' and end with `Rx'. The character ‘X’ in the string represents an interaction with an obstacle in the environment and could be either ‘R’ denoting ‘Reflection’, ‘D’ denoting ‘Diffraction’, ‘T’ denoting ‘Transmission' or `DS' denoting Diffused Scattering. The number of characters ‘X’ between the ‘Tx’ and ‘Rx’ in a given string represents the number of interactions, and the order from left to right of the characters denotes the sequence of interactions. Next, we strategically categorize MPCs into four groups according to the path model that is applicable to them. Observe in Table \ref{table_MPC_groups}, we have four MPC groups with their associated string that represent the interactions present for that group. As noted in section \ref{section_reflection_and_transmission}, transmissions do not affect the direction of propagation, hence for a given MPC, adding transmissions does not change the path model that applies. This is denoted by the substring `N$\times$-T-'.

\begin{table}[!htbp]
\caption{Four MPC groups categorized based on the Prop. Mech.}
\label{table_MPC_groups}
\centering
\footnotesize
\begin{tabular}{|c|c|c|c|}
\hline
 \textbf{MPC-1}& \textbf{MPC-2} & \textbf{MPC-3} & \textbf{MPC-4}   \\
\hline
\shortstack{Tx-Rx \\ Tx-(N$\times$-T-)-Rx}& \shortstack{Tx-R-Rx \\Tx-R-(N$\times$-T-)-Rx } & \shortstack{Tx-D-Rx \\ Tx-D-(N$\times$-T-)-Rx} &Others \\
\hline
\shortstack{Euclidean\\ path} &\shortstack{Euclidean\\ path}  &\shortstack{Diffraction\\ path}  & \shortstack{model\\ mismatch} \\
 \hline
\end{tabular}
\end{table}
Now for each MPC group across all receivers and transmitters, we calculate the probability of the FAP belonging to a particular MPC group ($P_{\text{FAP}}$ in \%) as
\begin{equation}
P_{\text{FAP}} = \sum_{\text{tx}} \frac{N_{\text{FAP},\text{tx}}\times 100}{N_{\text{rx}}}.    
\end{equation}
Here, $N_{\text{FAP},\text{tx}}$ is the number of receivers for which a particular MPC group is the FAP, $N_{\text{rx}}$ is the total number of receivers.

\section{Results And Discussion}

\subsection{First Arriving Path SNR \& Statistics}
\vspace{-1em}
\begin{figure}[!htbp]        \centering
        \includegraphics[clip, trim=3.5cm 9cm 4.5cm 9cm, width=0.8\linewidth]{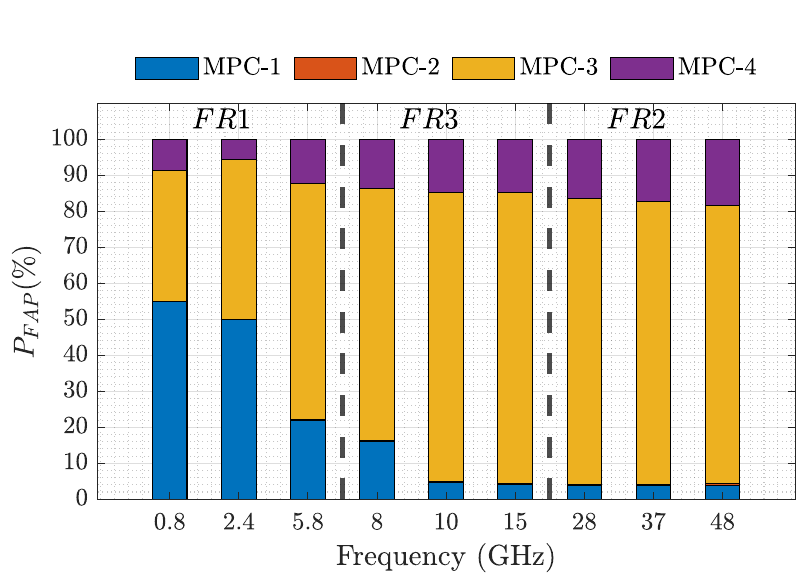}
    \caption{The SNR for the FAP across the FR1, FR2, and FR3 bands was evaluated using a FAP estimator with $T_{FAP} =20 dB$. For the FR2 and FR3 frequency bands, an additional transmission power of $10$ dBm was applied, while FR2 also included an extra $10$ dBm of receiver gain to compensate for the increased path loss. }
    \label{fig_P_FAP}
    \end{figure}
\begin{figure}[!htbp]
\centering
\includegraphics[clip, trim=3.5cm 9cm 4.5cm 9.5cm, width=0.75\linewidth]{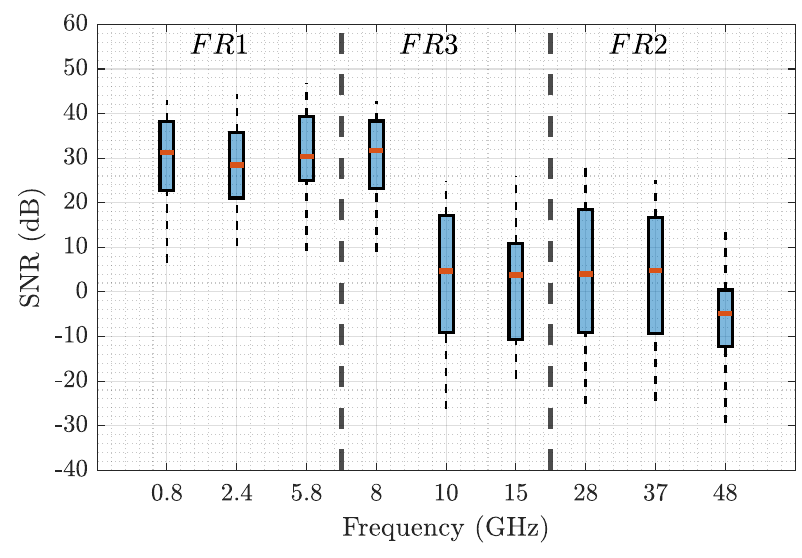}
\caption{Boxplot of received SNR versus frequency for the FAP, with a transmit power of 20 dBm for FR1 and 30 dBm for FR2 and FR3. Due to higher path loss, FR2 necessitates either additional receive-side processing gain or increased transmit power to maintain sufficient SNR.} 
    \label{fig_FAP_SNR}
\end{figure}
% The positioning performance is strongly influenced by the accuracy of our estimation of the FAP path length. This depends on two key factors: (a) the validity of our underlying assumption regarding the mathematical model governing the FAP path length, and (b) signal parameters like the FAP's SNR and the transmitted signal bandwidth. For instance, in the first scenario, if the FAP happens to be a diffraction path (MPC-3) and we employ LLS for positioning, it represents a case of model mismatch and results in reduced positioning performance. Thus, according to Table \ref{table_MPC_groups}, the suitable positioning technique, either LLS or D-NLS, should be chosen based on the dominant MPC group associated with the FAP. Next, even if we get the path model right, we should have sufficient SNR and signal bandwidth to enable proper estimation of the FAP length.
\par
In \figref{fig_P_FAP}, we illustrate the probability that the FAP belongs to different MPC groups for an FAP threshold of $T_{\text{FAP}}=20$ dB. The analysis is conducted for the FR1, FR2, and FR3 frequency bands across all receivers within the building. We influence the FAP estimator toward detecting shorter paths by increasing $T_{\rm FAP}$. However, increasing $T_{\text{FAP}}$ indefinitely is not feasible, as it is constrained by the influence of side lobes from nearby paths, including the strongest path, and the receiver's noise floor. At lower FR1 frequencies (below $5.8$ GHz), the FAP is predominantly composed of MPC-1 (transmission paths). This is because, as illustrated in \figref{fig_tansmissionLoss}, these frequencies exhibit significantly lower path loss compared to higher frequencies for concrete and drywall even after multiple transmissions. Further, we know that the Euclidean path represents the shortest path between two locations, and hence the FAP consists of MPC-1. 

In contrast from \figref{fig_tansmissionLoss} at FR2 and FR3 frequencies, transmission paths are considerably attenuated, and we expect the shortest signal propagation path to be the diffraction path according to Fermat's principle of least time \cite{duggal2025diffractionaidedwirelesspositioning}. For FR2 and higher FR3 frequencies we note that we can successfully isolate diffraction paths.  
Next, as shown in \figref{fig_FAP_SNR}, we evaluate the SNR associated with the FAP for these frequency bands with $T_{\text{FAP}}=20$ dB. Our goal was to achieve SNR $\ge0$dB for the FAP at majority of the receiver locations. Note, for FR1 frequencies we used a transmit power of $20$ dBm whereas for FR2 and FR3 frequencies we used $30$ dBm. Additonally, for FR2 frequencies we added $20$ dB of receiver processing gain to account for the extra path loss due to higher frequencies.

\subsection{3D Positioning Performance}

% \begin{minipage}
% \begin{figure*}[!htbp]
%     \begin{subfigure}{0.33\linewidth}
%         \centering
%         \includegraphics[clip, trim=3.5cm 9cm 4.5cm 9cm, width=1\linewidth]{figs/FAP_stat_BW_400_MHz_fap_threshold_0dB.pdf}
%     \caption{$T_{\text{FAP}}=0$ dB}
%     \end{subfigure}
%     \begin{subfigure}{0.33\linewidth}
%         \centering
%         \includegraphics[clip, trim=3.5cm 9cm 4.5cm 9cm, width=1\linewidth]{figs/FAP_stat_BW_400_MHz_fap_threshold_10dB.pdf}
%         \caption{$T_{\text{FAP}}=10$ dB}
%     \end{subfigure}
%     \begin{subfigure}{0.33\linewidth}
%         \centering
%         \includegraphics[clip, trim=3.5cm 9cm 4.5cm 9cm, width=1\linewidth]{figs/FAP_stat_BW_400_MHz_fap_threshold_20dB.pdf}
%         \caption{$T_{\text{FAP}}=20$ dB}
%     \end{subfigure}
%      \caption{The FAP estimator fails to isolate diffraction paths (MPC-3) ($P_\text{FAP}$ is low for MPC-3) for (a) FR1 band. However, it successfully isolates diffraction paths ($P_\text{FAP}$ is high for MPC-3 for $T_{\text{FAP}}=10dB,20dB$) for (b) FR3 band (c) ($P_\text{FAP}$ is high for MPC-3 for $T_{\text{FAP}}=20dB$) for FR2 band. }
%     \label{fig_P_FAP}
% \end{figure*}
\begin{figure*}[!htbp]
    \begin{subfigure}{0.33\linewidth}
        \centering
\includegraphics[clip, trim=0cm 0cm 0cm 0cm, width=1\linewidth]{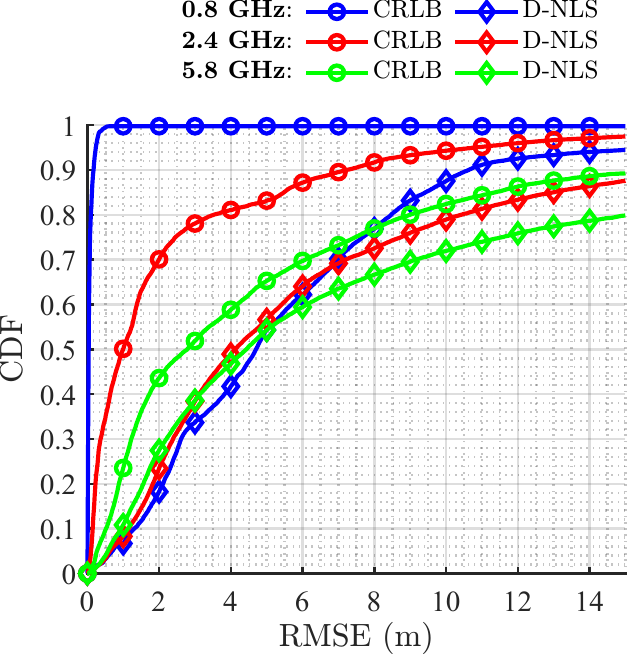}
        \caption{FR1 band, Tx power:20 dBm}
        \label{fig_3D_positioning_FR1}
    \end{subfigure}
    \begin{subfigure}{0.33\linewidth}
        \centering
\includegraphics[clip, trim=0cm 0cm 0cm 0cm, width=1\linewidth]{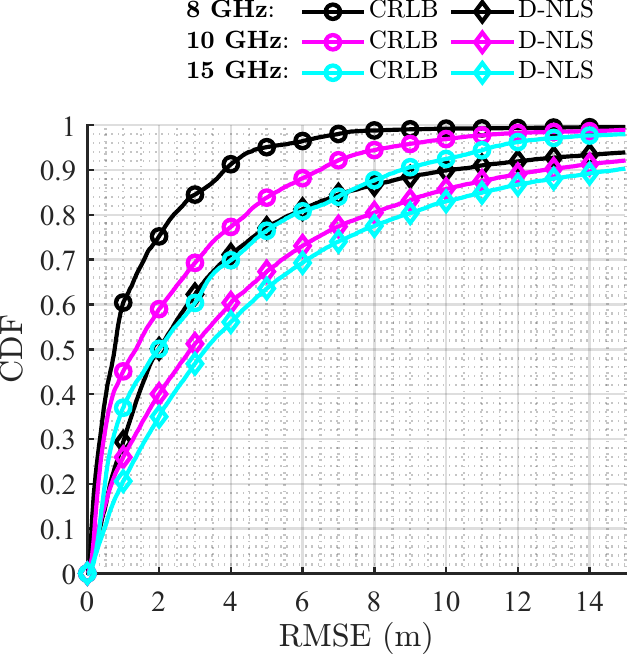}
        \caption{FR3 band, Tx power:30 dBm}
        \label{fig_3D_positioning_FR3}
    \end{subfigure}
        \begin{subfigure}{0.33\linewidth}
        \centering
\includegraphics[clip, trim=0cm 0cm 0cm 0cm, width=1\linewidth]{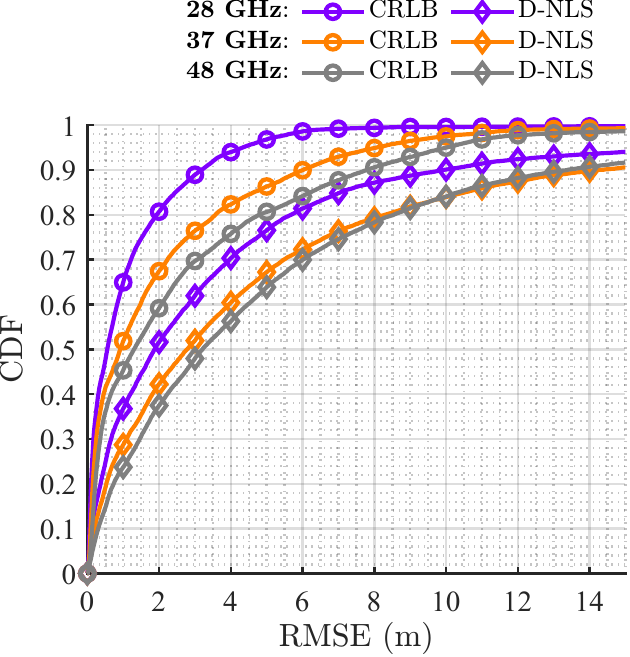}
        \caption{FR2 band, Tx power:30 dBm}
        \label{fig_3D_positioning_FR2}
    \end{subfigure}
\caption{3D positioning performance for FR1, FR2 and FR3 bands for $T_{\rm FAP}=20$dB and Bandwidth B$=400$ MHz. } 
    \label{fig_positioning_results}
\end{figure*}
% \end{minipage}
For each receiver-transmitter pair, the shortest diffraction path (MPC-3) is estimated at the receiver location and its SNR is used in the CRLB calculation to determine the lower bound of the 3D RMSE for the position estimate. This assumes perfect isolation of the diffraction MPC at all receiver locations and illustrates the lower bound to the possible positioning performance using the diffraction paths.
Next, for each receiver location we estimate the FAP based on section \ref{section_first_arriving_path_estimator}. This is used to derive 3D position estimates using D-NLS and LLS positioning algorithms. A model mismatch occurs if D-NLS is applied when the FAP does not belong to MPC-3, or if LLS is applied when the FAP does not belong to MPC-1. In such cases, contrary to expectations, positioning performance will not improve with increasing SNR or bandwidth. Observe in \figref{fig_3D_positioning_FR1} that for frequencies less than $5.8$ GHz, both D-NLS and LLS are very far from CRLB. Furthermore, LLS outperforms D-NLS. This is because, as shown in \figref{fig_P_FAP}, even with an increase in $T_{\text{FAP}}$, the FAP remains similarly distributed between MPC-1 and MPC-3. Physically, the transmission paths (MPC-1) are not sufficiently attenuated at these frequencies. Consequently, a model mismatch occurs at a significant number of receiver locations with both positioning algorithms.
\par On the other hand, observe in the FR2 band in \figref{fig_3D_positioning_FR2}, D-NLS significantly outperforms LLS, indicating that most FAPs are diffraction paths (MPC-3) with sufficient SNR. However, this requires increased transmit power and receiver processing gain to compensate for the significant path loss. 
\par The sweet spot in 3D positioning performance is achieved at frequencies approximately within the FR3 band in \figref{fig_3D_positioning_FR3}. At these frequencies, the transmission paths experience sufficient attenuation, enabling effective isolation of diffraction paths using the first-arriving path principle. Additionally, compared to the FR2 band, these frequencies exhibit lower path loss, reducing the need for increased receiver processing gain to attain the desired SNR.

\section{Conclusion}
% \GD{Summarize results as a big picture. Future research will focus on identifying and isolating both diffraction and transmission paths (such as at FR1 frequencies) and then applying the appropriate mathematical path model to further enhance position estimation accuracy.}
We analyze diffraction from the edges of the windows as a key propagation mechanism for frequencies that span the upper FR1, FR2 and FR3 bands. Using ray-tracing simulations, we modeled transmission, reflection, diffraction, and diffuse scattering in an O2I scenario aimed at estimating the position of an indoor receiver with outdoor UAV transmitters. Our results demonstrate that diffraction paths can be isolated using the first-arriving path principle for frequencies in the upper FR1 to FR2 range including FR3. This is because at these frequencies transmission paths are significantly attenuated, increasing the likelihood that the next shortest path is a diffraction path. These diffraction paths carry crucial position information and help mitigate NLoS bias in 3D position estimation.
 However, for diffraction-based NLoS positioning to perform effectively at FR2 frequencies, the additional path loss compared to FR1 \& FR3 must be compensated by high receiver processing gain or increased transmit power. Future research will focus on identifying and isolating both diffraction and transmission paths (such as at lower FR1 frequencies) and then applying the appropriate mathematical path model to further enhance position estimation accuracy.    

\section*{Acknowledgment}
We sincerely appreciate the support provided by NIST PSCR PSIAP (70NANB22H070), NSF (CNS-1923807, CNS-2107276), and the NIJ Graduate Research Fellowship (15PNIJ-23-GG-01949-RES). Furthermore, we express our gratitude to Tarun Chawla of Remcom for granting access to Wireless InSite \cite{wirelessinsight}.

% \AK{In this paper, we first discuss the EM principles that govern the propagation of signals through building materials and its dependence on signal frequency. We then apply a diffraction based path model derived from first principles to perform NLoS positioning. We make use of we use ray-tracing simulations to analyze $P_{FAP}$ statistics and demonstrate that receivers using the FAP principle for positioning predominantly pick the diffraction (MPC-2) path as the first arriving path for NLoS positioning in upper FR1, FR2 and FR3 frequencies. We note that at lower frequencies the receiver picks the euclidean path for positioning leading to a model mismatch that degrades positioning performance using D-NLS algorithm while we observe best performance in the proposed FR3 bands due to sufficient attenuation of transmission paths resulting in reduced model mismatch while demonstrating lower path loss compared to FR2 bands resulting in better NLoS positioning performance. }

\bibliography{refs}
\bibliographystyle{IEEEtran}
\end{document}